\documentclass[12pt]{iopart}
\usepackage{graphicx}
\begin{document}
\title[Clean and adsorbate-covered decagonal Al-Co-Ni]{Bulk and surface structure of the clean and adsorbate-covered decagonal Al-Co-Ni quasicrystal}
\author{S Burkardt$^{1}$, S Deloudi$^{1}$, M Erbudak$^{1,2}$, A R Kortan$^{3}$,\\ M Mungan$^{2,4}$, W. Steurer$^{1}$}
\address{$^1$ETH Zurich, 8093 Zurich, Switzerland}
\address{$^2$Bo\u gazi\c ci University, 34342 Bebek, Istanbul, Turkey}
\address{$^3$56 Christy Drive, Warren, 07059 NJ, USA}
\address{$^4$The Feza G\"ursey Institute, P.O.B. 6, \c Cengelk\"oy, 34680 Istanbul, Turkey}
\ead{erbudak@phys.ethz.ch}

\date{\today}

\begin{abstract}
We review our Al adsorption experiments on the tenfold-symmetry surface of the decagonal Al-Co-Ni quasicrystal and present computational simulations of adsorption on a structural model based on a fundamental Al-Co cluster with 20\,\AA\ diameter, symmetry $\overline{10}2m$, and 8\,\AA\ periodicity. This cluster is the building unit of $\tau ^2$-Al$_{13}$Co$_4$, from which, by a sequence of minor changes, the structures of the phases in the stability region of decagonal Al-Co-Ni can be derived. The model used for the decagonal Al$_{70}$Co$_{15}$Ni$_{15}$ is an idealized model with a two-layer periodicity (4 \AA) and no chemical or structural disorder. We find that the bulk and surface properties of this model are in good agreement with experiments. Our molecular-dynamics simulations of Al adsorption reproduce the experimental results and  show that by varying the thermal relaxation rates of the adsorbed layer, a variety of different surface morphologies can be achieved. We also present our recent experiments on dissociative adsorption of oxygen on the decagonal surface. 
\end{abstract}

\maketitle

\section {Introduction}

One may expect that unusual structures give rise to extraordinary properties. In the case of low-dimensional systems, these extraordinary properties substantially contributed to the high-impact field of nanotechnology. For quasicrystals, comparably exciting applications have not yet been realized. Still, several hundred published reports indicate an ongoing scientific curiosity about bulk and surface structure, and properties of quasicrystals. Comprehensive reviews of the field have recently been published \cite {sharma,Steurer}.

An interface formed by two materials with dissimilar atomic structures is called hetero-epitaxy and cannot maintain conditions for epitaxial growth on a global scale. Locally, however, commensurability may persist on nanometer scale leading to the 
formation of self-size-selected monocrystalline domains \cite {Arthur}. Such nanometric configurations may find under suitable electronic conditions applications as low-dimensional devices. A good candidate is the interface where an ordinary crystal and a quasicrystal meet. Preparation methods for the generation of structurally well-defined, clean quasicrystalline surfaces have been established which make quasicrystals interesting substrates for thin film growth. In particular, the surface perpendicular to the tenfold-symmetry axis of the decagonal Al-Co-Ni is very suitable as a substrate for epitaxial growth due to its flatness. Up to now, several crystal-quasicrystal interfaces have been investigated \cite{sharma,Thiel}.

Here, we investigate first the structure of Al-Co-Ni (Al$_{70}$Co$_{15}$Ni$_{15}$), which is quasiperiodic in two dimensions, periodic in the third dimension, and has decagonal diffraction symmetry. Investigations of its bulk structure have revealed a columnar prismatic morphology with the column axis parallel to the periodic direction \cite{refik2,steurer2}. We then review our experimental and numerical work on the adsorption of Al atoms on decagonal Al-Co-Ni and present novel experimental results for the dissociative chemisorption of oxygen on the same surface. 

In previous work, we have studied numerically the effect of the relative strength of  mutual interactions of adsorbed atoms (adatoms) with respect to their  interactions 
with the substrate atoms \cite{bilki,mungan} on the structure of the forming adsorbate. Noting that deposition is a non-equilibrium process where the ratio of the  thermal relaxation to the deposition rate of the adatoms has influence on  the resulting morphology of the adsorbate \cite{barabasi}, we will describe how the adsorbate layer (adlayer) structure depends on the rate at 
which adatoms thermally equilibrate with the substrate.

The paper is organized as follows: In Section II we outline the experimental details. Section III focuses on the structural model of dedacagonal Al-Co-Ni and its evaluation by comparing single site scattering calculations (SSC) with experimental secondary electron imaging (SEI) patterns, as well as surface diffraction calculations with experimental low-energy electron diffraction (LEED) patterns. We next turn to adsorption on the decagonal Al-Co-Ni surface in Section IV. We first focus on our experimental and numerical work on Al deposition on the decagonal surface in Section IV A. Next, we present experimental results on the dissociative oxygen adsorption on the decagonal surface in Section IV B that occurs during the initial stages of oxidation. Section V summarizes our results 

\section {Experimental}

In this report we have used an Al$_{70}$Co$_{15}$Ni$_{15}$ sample with dimensions 5\hspace{1pt}$\times$\hspace{1pt}3\hspace{1pt}$\times$\hspace{1pt}1\,mm$^3$ which was oriented by means of the x-ray Laue method with an accuracy of $\pm$\,0.5$^\circ$ along the decagonal symmetry axis, cut, and polished mechanically to an optical finish \cite{Kortan}. The sample surfaces were cleaned in ultrahigh vacuum in the lower 10$^{-10}$-mbar region by cycles of sputtering with Ar$^+$ ions (1.5\,keV, 4.5\hspace{1pt}$\times$\hspace{1pt}10$^{-7}$\,A/mm$^2$) at 680\,K and heat treatment at 900\,K for 30 minutes. The temperature was measured by means of a chromel-alumel (K-type) thermocouple pressed onto the sample surface. A three-grid back-view display-type low-energy electron diffraction (LEED) system, operating at low microampere primary currents, had a total opening angle of about 100$^{\circ}$. Thus, a momentum transfer of 2.78\,\AA$^{-1}$ could be detected at 50\,eV at normal primary-electron incidence. The diffraction patterns were recorded by a 16-bit charge-coupled device camera. The sample preparation was monitored by the quality of the LEED pattern and by inspecting the scans of Auger electron spectroscopy (AES). The latter were recorded using a cylindrical mirror analyzer operating with a constant relative resolution of $\Delta E/E = 0.8\,$\%. LEED patterns are registered after the sample was cooled down to room temperature (RT). The position and the size of the Bragg spots in LEED observations are used to extract real-space information about the atomic structure, the size, and the orientation of the surface textures. Relevant procedures can be found in the literature \cite{LEED1,LEED2,Rouven}. Experimental details have previously been reported \cite{review,Thomas,Longchamp1}.

SEI experiments are performed using the LEED display system under the same sample position, while primary-electron energy is increased to 2\,keV. Primary electrons are used to excite secondary electrons in a surface region the depth of which is determined by the mean free path of electrons at the corresponding energies. In SEI, the secondary electrons generated by a source atom, located at a lattice site, are focussed in the forward-scattering direction by scattering at the neighboring atoms and thus their intensities are enhanced along interatomic directions. The secondary electrons, emitted into vacuum and detected by the spherical collector, thus represent a real-space central projection of the local symmetry around the source atom, with many sources contributing incoherently to the pattern. Therefore, SEI helps us observe the local symmetry properties of atoms in a near-surface region averaged over a volume. 
The latter is defined by the area of the spot of the primary electrons and the inelastic  mean free path of the secondary electrons at the electron energy used in the experiment \cite{review}.  Besides revealing dense atomic rows at and below the surface, secondary-electron patterns also display so-called Kikuchi bands, which originate from Bragg diffraction of quasielastically scattered electrons at parallel planes of high atomic density \cite{baba}.

\section{Structural Model of the Quasicrystal and its Evaluation}
\label{Sofia}

\subsection{The Model}

An idealized model of the decagonal Al$_{70}$Co$_{15}$Ni$_{15}$ quasicrystal (type-II phase) has been used for the simulations of the adatom deposition on the tenfold-symmetry surface. This model is part of a systematic and consistent cluster-based modeling of the different modifications of decagonal Al-Co-Ni and its approximants. The modeling shows good agreement with available experimental data based on x-ray diffraction and HAADF-STEM (high angle annular dark field-scanning transmission electron microscopy) images. A detailed discussion of the models and their derivation can be found elsewhere \cite{sofia10}. In the following, a short overview of the deduction path and the model of Al$_{70}$Co$_{15}$Ni$_{15}$  will be given.

A fundamental Al-Co cluster has been proposed as the building unit of $\tau ^2$-Al$_{13}$Co$_4$ and as a starting point for the derivation of structure models for the  decagonal phases. This choice of the fundamental cluster enables us to model the different modifications on a common basis and offers thus a rational approach to elucidate the complex stability field of decagonal Al-Co-Ni. The Al-Co cluster has $\sim$\,20\,\AA\ diameter, $\bar1\bar02m$ symmetry, and $\sim$\,8\,\AA\ translation period. It consists of two flat layers located at the mirror planes at z = 0 and 1/2 (the cluster axis is denoted z-axis in the following), as well as of two puckered layers at z = 1/4 and 3/4, respectively. The mirror planes perpendicular to the $\bar1\bar0$ axis relate the puckered layers to each other. The structure of the $\tau ^2$-Al$_{13}$Co$_4$ approximant is obtained by decorating the proper periodic tiling with the fundamental cluster.

\begin{figure}[t]
\includegraphics[width=1.0\columnwidth]{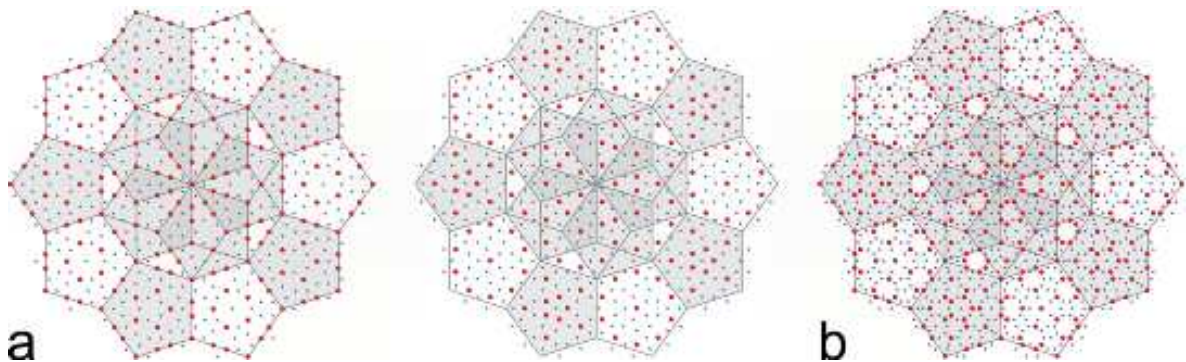}
\caption{The two layers of the supercluster building the model for the decagonal Al$_{70}$Co$_{15}$Ni$_{15}$ are shown in (a). The supercluster is built by two subunits, cluster 1 and 2, which are highlighted by non-filled and gray filled pentagons. The projection of the supercluster along the periodic direction is shown in (b).}
\label{fig:super2}
\end{figure}

The decagonal phases and the W-approximant are described by two clusters, further referred to as cluster 1 and cluster 2. Due to the close structural interconnection of the decagonal phases with their rational approximants, both clusters can be derived from the fundamental cluster through small structural changes. Those changes also include the introduction of chemical disorder, which is considered to be an intrinsic structural feature of the decagonal phases and the W-approximant. The structures are then obtained by decorating appropriate (quasiperiodic) tilings with the corresponding clusters. The structure of the type-II phase is obtained by decorating a rhombic Penrose tiling with a `supercluster' (diameter $\sim$\,80\,\AA) formed by cluster 1 and 2. Overlapping of the clusters is allowed only in certain ways, defined by the underlying tiling. Here, flip-positions and mixed occupations (chemical disorder) are generated, which are in good agreement with experimental data (for more details see Ref. \cite{sofia10}). The four-layer (8\,\AA) periodicity of the columnar clusters is solely due to puckering in the periodic direction, which means that the layers located at the mirror planes are identical. In our calculations, we used an idealized model consisting only of flat layers, and thus with a two-layer (4\,\AA) periodicity. We furthermore simplified the model by not allowing partial occupation of flip-positions or chemical disorder. Fig.~\ref{fig:super2} shows the structure of the two layers of the supercluster as used, as well as its projection along the periodic direction.

\subsection {Single-Site Scattering Calculations}

In a simplified picture, a direct projection imaging of the vectors connecting the next-neighbor atoms with a reference atom is responsible for the formation of the SEI patterns. In this case,  a computational interpretation of the pattern is straightforward using a simple geometric analysis. One neglects the wave-mechanical character of the electrons and considers the forward focusing of the electrons along main crystallographic directions for small emitter-scatterer distances in order to reproduce the main symmetry features of the experimental pattern \cite{eskiPRB,cham}.

A more realistic simulation of the SEI patterns can be achieved by quantum-mechanical SSC calculations, which are developed to simulate the angular distribution of the intensity of photoemitted electrons \cite{fadley}. In our simulations of the secondary-electron patterns, we made two assumptions: Firstly, we neglected the forward scattering of the primary-electron beam. This is justified, since the geometry is fixed during the experiment and the forward scattering of the primary electrons has an effect mainly on the absolute intensities in the SEI patterns. Secondly, the quasielastically backscattered electrons are described as photoelectrons emitted from an {\it s} level. This approximation may not critically affect the calculated patterns, because it has already been shown that the main features of photoelectrons measured for different core levels are the same \cite{cham}. In the calculations, scattering of secondary electrons in a cluster is described by means of phase shifts derived from the muffin-tin potentials of Al and Co \cite{clementi}. Both in the structural model and the SSC calculations we use Co atoms for the transition metals.   

The present calculations for the tenfold-symmetry surface of a decagonal Al-Co-Ni quasicrystal are based on a cluster with 513 atoms located in a cone with the axis parallel to the decagonal direction of the crystal with a total opening angle of $110^{\circ}$. The base area of the cone is the decagonal surface. We have chosen 77 atoms near the apex of the cone as sources of secondary electrons. We have constructed a similar cone with the axis normal to the tenfold-symmetry direction of the quasicrystal. The scattering intensity within an opening angle of $100^{\circ}$ is plotted in such a way as to furnish direct comparison with the central-projection imaging of the SEI experiments.

\begin{figure}[!t]
\includegraphics[width=0.9\columnwidth]{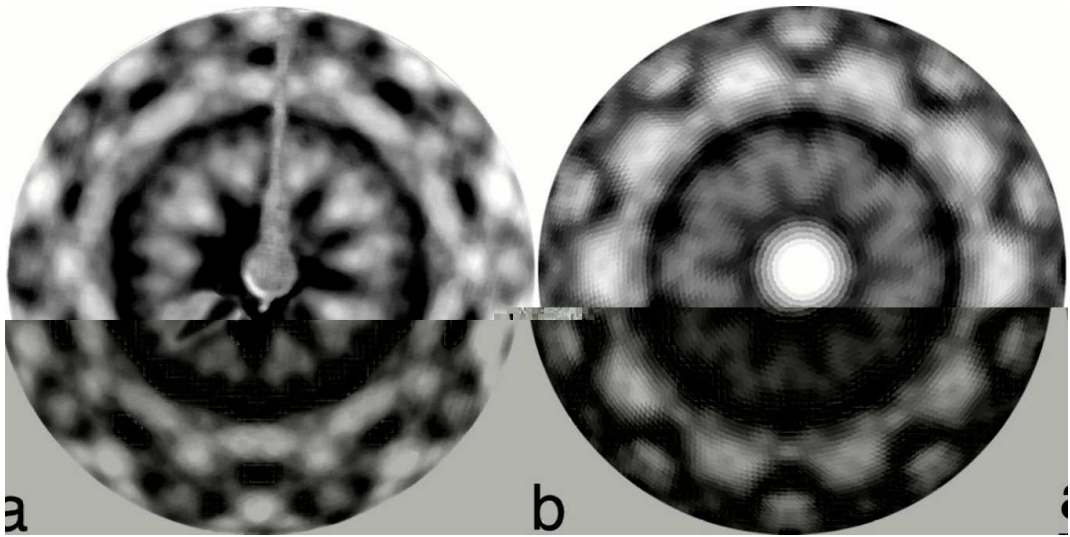} 
\caption{The SEI pattern (a) from the tenfold-symmetry surface of the decagonal Al-Co-Ni  obtained at 2\,keV is compared with the results of the SSC calculation (b), performed using the model oriented appropriately.}
\label{fig:onlusei}
\end{figure}

Fig.~\ref{fig:onlusei}a presents an SEI pattern from the Al-Co-Ni sample for which the tenfold-symmetry axis almost coincides with the surface normal \cite{eskiPRB,busch}. This was also the incident direction of the 2\,keV primary electrons. Scanning the electron beam across the sample surface left the pattern unchanged confirming that the specimen consisted of one single grain. The overall decagonal symmetry is the most striking feature observed in the pattern. Significant contribution to the pattern originates from Kikuchi bands due to planes inclined by $\theta = 30 \pm 1^{\circ}$ from the decagonal direction. Bright patches of increased electron intensity appear on concentric rings at different polar angles. Features on the innermost ring occur at $\theta = 18 \pm 1^{\circ}$. A concentration of bright spots, which are arranged in decagonal symmetry, is located at $\theta = 35 \pm 1^{\circ}$. Between these two rings of bright patches, a dark circular area with no significant electron intensity is observed at $\theta = 25 \pm 1^{\circ}$ suggesting a tube-like hollow structure along the decagonal axis, confirming earlier results about columnar channels \cite{refik2,steurer2}. Near the edge of the pattern, additional bright patches are observed which are distributed also in decagonal symmetry.

\begin{figure}[!t]
\includegraphics[width=1.0\columnwidth]{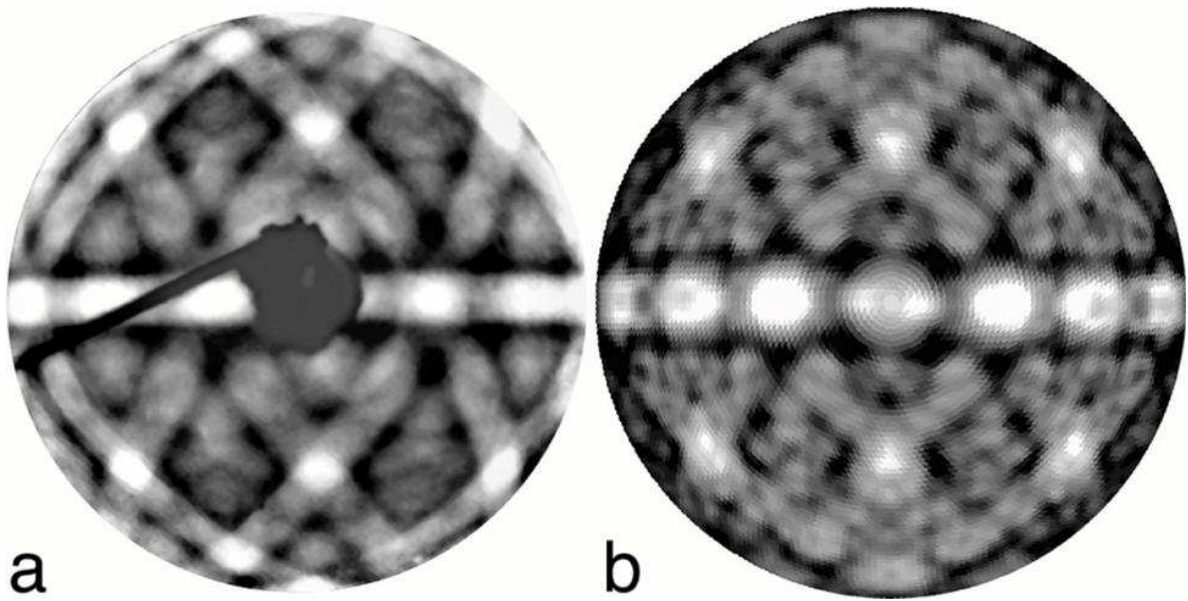} 
\caption{The SEI pattern (a) from the twofold-symmetry surface of the decagonal Al-Co-Ni  obtained at 2\,keV is compared with the result of the SSC calculation (b) performed with the coordinates of the model, oriented as described in the text.}
\label{fig:ikilisei}
\end{figure}

Fig.~\ref{fig:onlusei}b shows the result of the SSC calculations on the model structure. It is evident that the pattern is tenfold symmetric. Major emission features are located on two rings with the polar angle of $34^{\circ}$ and $46^{\circ}$ each consisting of ten patches. There is a good agreement between these features and the experimental ones presented in Fig.~\ref{fig:onlusei}a. The width of calculated intensity maxima is usually larger than the experimental bright patches, because single-scattering calculations overestimate intensities along close-packed directions. Multiple scattering would produce some defocusing, which is most effective for short emitter-to-scatterer distances, and would increase the degree of agreement between experiments and calculations \cite {zheng}.

Fig.~\ref{fig:ikilisei} presents analogous results obtained from the twofold-symmetry surface. The characteristic feature of the experimental pattern is the bright band oriented horizontally and coinciding with the equatorial plane of the scattering geometry, thus forming a mirror-symmetry plane of the pattern. This band originates from tenfold-symmetry planes oriented perpendicular to the surface. Within the decagonal plane, we distinguish patches at every $18^{\circ}$, representing alternately two kinds of twofold-symmetry axes of the quasicrystalline structure, denoted A2D and A2P \cite{qin}. We note that A2D coincides with the primary-electron incidence within a few degrees. Another significant contribution to the pattern are the bright Kikuchi bands that cross pairwise the bright band due to the twofold-symmetry plane at A2P axes. We  note that they form a series of equilateral triangles that have one side, spanning $36^{\circ}$, in common with the decagonal plane \cite{eskiPRB,tommy}. Hence, a polar rotation of the sample by multiples of $36^{\circ}$ around an axis normal to the tenfold-symmetry axis aligns another A2D axis with the display axis, and the resulting SEI pattern looks exactly the same as that shown in Fig.~\ref{fig:ikilisei}a. This observation confirms the overall tenfold-symmetry property of all the twofold-symmetry planes together, located within the volume analyzed in SEI.

Fig.~\ref{fig:ikilisei}b depicts the SSC calculations from our cluster of atoms aligned in order to expose the twofold-symmetry surface. Electron intensities along some interatomic directions are over estimated, in particular those lying in the twofold-symmetry planes and the band-like emission is underestimated as a result of the single-scattering process. Other than that, all the symmetry features  in the experiment and those computed for the Al-Co-Ni structural model agree well.

\subsection {Surface Structure}

\begin{figure}[!t]
\includegraphics[width=0.9\columnwidth]{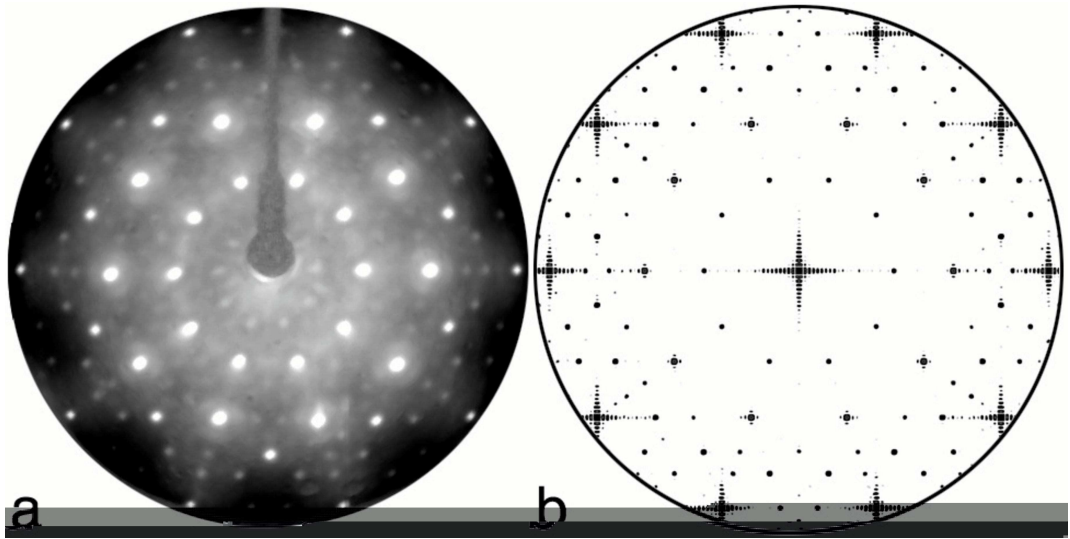} 
\caption{Patterns of (a) electron diffraction from the tenfold-symmetry surface of the decagonal Al-Co-Ni at a primary-electron energy of 50\,eV compared with (b) the Fourier transform of atomic coordinates of the same surface. Note that only one single layer is considered.}
\label{fig:leedonlu}
\end{figure}

Fig.~\ref{fig:leedonlu}a shows a LEED pattern from the tenfold-symmetry surface of the decagonal Al-Co-Ni at a primary-electron energy of 50\,eV. The tenfold rotational symmetry of the spot distribution and the relatively low background intensity indicate that the surface has good structural property. There have been efforts to extract real-space information about the atomic structure and chemistry at and below the surface by using the energy dependence of a chosen number of diffraction-spot intensities \cite{ferralis}. Here, we simply compare the positions of diffraction spots at a single energy with model coordinates, which have been transformed into reciprocal space.

Fig.~\ref{fig:leedonlu}b illustrates the calculated intensity distribution in reciprocal space obtained from one of the surfaces of the bilayer quasicrystal. We calculated the intensity pattern using a simple Fourier transform, thereby ignoring any effects arising from multiple scattering. Atom-specific scattering factors have been ignored as well. The orientation and scale of the calculated pattern are chosen to match those of the experimental LEED spots. Within these approximations the diffraction patterns calculated from either of the two surfaces yield identical spot locations that agree well with the LEED results. A calculation taking into account multiple scattering and atomic specificity would also yield varying spot intensities and therefore slightly alter the calculated image seen in Fig.~\ref{fig:leedonlu}b.

\section {Adsorption on the tenfold-symmetry surface}

We describe two deposition experiments with the tenfold-symmetry surface of Al-Co-Ni: In the first experiment, we vacuum deposit Al onto a clean decagonal surface and observe the growth of several Al domains in the native face-centered cubic (fcc) structure, each aligned with the (111) surface parallel to the decagonal face and in equal azimuthal increments of $36^{\circ}$ producing a pseudo-decagonal symmetry. In order to account for some of the details of the growth process qualitatively, we perform molecular-dynamics simulations of the deposition of adatoms on a quasicrystalline rigid bilayer. The simulation of multilayer growth confirms the formation of local fcc domains with their (111) faces parallel to the decagonal surface of the substrate. The azimuthal alignment of the domains is in fivefold registry with the quasicrystal surface. The morphology of the growing adlayer varies from cluster-type growth to homogenous growth depending on parameters such as the thermal equilibration time of the adlayer and the relative strength of the adatom interactions. In the second experiment, we introduce pure oxygen onto the quasicrystalline surface at elevated temperatures. Similar to Al(111), we observe the dissociative adsorption of oxygen in the form of a chemisorbed, but nevertheless well-ordered, hexagonal domains at very low coverages, indicating that the binding between the adsorbate atoms dominates over the oxygen-quasicrystal interaction.

\subsection {Aluminum Adsorption}
\subsubsection{The Experiment}

\begin{figure}[!t]
\includegraphics[width=1.0\columnwidth]{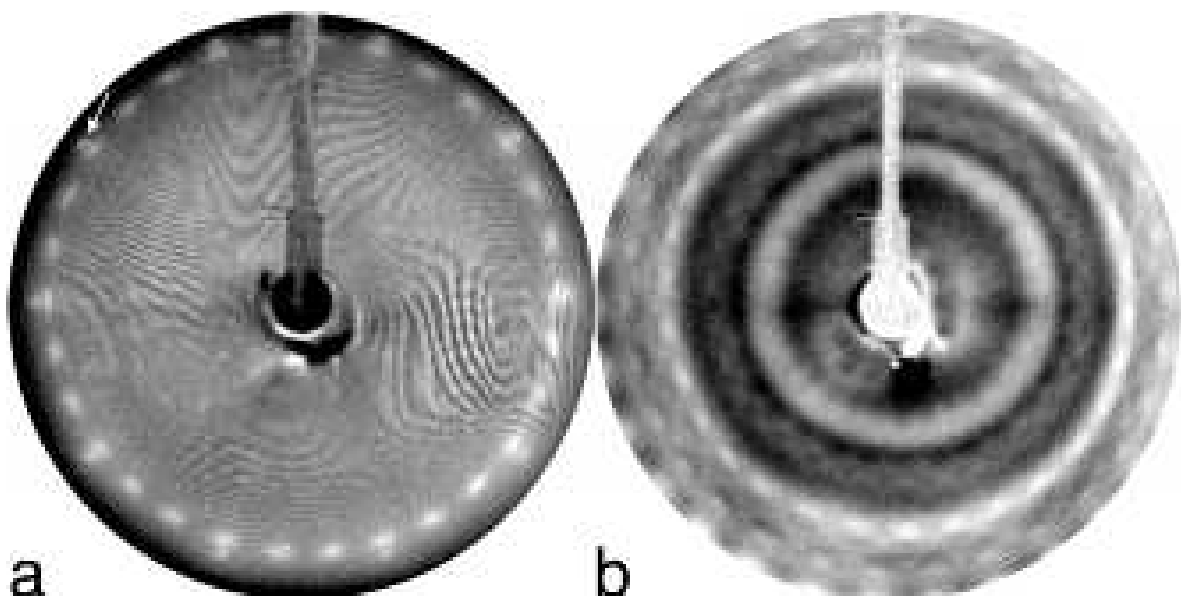} 
\caption{LEED (a) and SEI (b) pattern from the tenfold-symmetry surface of decagonal {Al-Co-Ni} after Al-deposition. The real-space SEI pattern contains intensity patches oriented in thirtyfold-symmetry on rings of constant polar angles, confirming the existence of ten sixfold-symmetric crystalline surface structures.}
\label{fig:Alevap}
\end{figure}

Al evaporation was done with a water-cooled and power-regulated effusion cell. The evaporation rate was $0.8\pm0.1$\,\AA/min, calibrated, in a separate experiment, using the Al $L_{2,3}VV$ Auger transition signal during evaporation on a Cu sample. The absolute calibration of the flux from the atom source has been previously done by comparing the changes in the mass of a quartz microbalance for a given condition of operation of the source \cite {eski}. During the Al deposition the substrate  was maintained at RT. 
LEED and SEI observations show that Al vacuum deposited onto the decagonal surface of Al-Co-Ni forms nanometer-sized crystallites starting at a few monolayers that remain stable up to a thickness of 120\,\AA\  \cite{Thomas}. Low Al coverages mostly reproduce quasicrystalline spots without a clear evidence of structure due to Al overlayer. This observation points to an Al structure, which is either amorphous or quasicrystalline owing to pseudomorphic growth. Fig.~\ref{fig:Alevap}a displays a LEED pattern recorded at $E_p=55$\,eV after deposition of 6\,\AA\ Al. At this coverage, however, most of the substrate contribution is faded away, except for the ten spots discernible near the rim of the screen and distributed in equal azimuth location. One such spot is marked by the white arrow. We observe that Al deposition also leads to the formation of azimuthally elongated diffraction spots, all lying on a circle with a polar angle of 45.0\,$\pm 1^{\circ}$. An analysis of the spot positions and spot profiles reveals that Al forms the native fcc structure exposing  the (111) face. Each diffraction patch actually consists of two spots separated azimuthally by about $2.5^{\circ}$, and the domain size is $30-35$\,\AA\ as extracted from spot-profile analysis \cite{Thomas}. The  diffraction spots from the Al-Co-Ni substrate persist up to an Al thickness of 12\,\AA\ suggesting a cluster growth but not a homogenous growth.

We note the orientational coincidence of the tenfold-symmetric pattern from the substrate with the thirty-fold adlayer pattern indicating that the surface layer is in azimuthal registry with the quasicrystalline substrate. The fact that the adsorbed layer azimuthally locks to the  substrate proves that the underlying quasicrystal serves as a structural template for the Al film growth. Epitaxial growth conditions, like lattice and chemical matching that are observed in crystals, are apparently satisfied on a local scale. The LEED pattern for the aluminum adsorption, shown in Fig.~\ref{fig:Alevap}a, is similar to that resulting from oxygen adsorption, as depicted in Fig.~\ref{fig:Leedox}. In both cases a thirty-spot diffraction pattern attributed to the adsorbate layer is seen, however, the relative azimuthal registry of the adsorbates is offset by $6^{\circ}$. 

Time-reversal symmetry in LEED experiments renders the distinction between threefold and sixfold symmetric surface structures difficult; both result in a sixfold-symmetric diffraction pattern \cite{reversal,reversal2}. Likewise, by inspecting the LEED pattern shown in Fig.~\ref{fig:Alevap}a, we cannot distinguish whether there are five or ten sets of Al domains on the quasicrystal surface without an energy-dependent intensity analysis of the diffraction spots. This ambiguity can be resolved by inspecting the real-space SEI pattern. Fig.~\ref{fig:Alevap}b presents an SEI pattern obtained from an Al layer of 120\,\AA. Secondary electrons leaving the sample into vacuum at polar angles of $20^{\circ}$ and $35^{\circ}$ are concentrated in two diffuse bright circles. Near the rim we further recognize two circles of 30 discrete patches. A single domain fcc (111) surface structure would lead to three spots at the polar angle of $35.3^{\circ}$ due to the $\langle110\rangle$ directions \cite{review}. Since 30 bright spots are observed on  rings with distinct polar angles, two sets of ten Al(111) domains must be present on the surface.

The combination of LEED and SEI shows that Al crystallizes in two sets of domains, each about 32\,\AA\ in size, in the native fcc structure with the (111) face aligned parallel to the decagonal surface of the substrate. Within each set, the domains are rotated by $36^{\circ}$ increments, as a consequence of either fivefold-symmetric terraces each inverted by $180^{\circ}$ or a local tenfold symmetry of the substrate. For the structural model presented here, only the first possibility is feasible. The two sets are displaced by about  $2.5^{\circ}$ relative to each other  leading to the azimuthal elongation of the diffraction spots.

\subsubsection{Calculations}

Energy calculations of the Al-quasicrystal interface demonstrated the existence of favored orientations of a single hexagonal layer of rigid adatoms on the decagonal quasicrystal surface \cite{Thomas}. The results lead to a qualitative explanation of the experimentally observed adlayer domains, their orientation, and sizes.  The success of the rigid adlayer lattice assumption, prompted us to further investigate the extent of its validity by performing molecular dynamics simulation of the adatoms on a quasicrystalline substrate. 
We investigated the three-dimensional adlayer growth by carrying out extensive numerical simulations of the deposition of 3.5 monolayers (ML) corresponding to 11\,000 adatoms. As the substrate, we used a 6\,000-atoms section with dimensions  $150\times150\times 2.05$\,\AA $^3$ of the model of the decagonal Al$_{70}$Co$_{15}$Ni$_{15}$ quasicrystal, described in Section \ref{Sofia}. The large lateral size of the unit cell permits us to observe the formation of multiple domains and reduces artefacts generated by boundary conditions. Here we consider the deposition process under conditions of very slow energy dissipation, leading to a large thermal relaxation time for impinging adatoms.

The adatoms are deposited by injecting them at an initial height $z_0$ above the substrate surface with a uniform initial velocity $v_0$. We assume that the substrate remains rigid and use a cooling mechanism to maintain the adlayer at a constant temperature $T$, as we will explain below. The interaction between the adatoms, as well as between the adatoms and the substrate atoms is assumed to be pairwise and of Lennard-Jones type. The characteristic energy and length scales in the computations are given by $\epsilon$ and $\sigma$, respectively.  This type of interaction was chosen for simplicity, since the mechanism of domain formation is fairly generic and thus is not expected to be critically sensitive to the actual details of the underlying interactions. This is also corroborated by the similar adsorption of Xe on the same surface \cite{setya2}.   

Letting ${\bf r_\alpha}$ and ${\bf r_i}$ denote the coordinates of a substrate atom $\alpha$ and an adatom $i$, respectively, 
we numerically evaluate the equations of motion 
\begin{eqnarray}
m \ddot{\bf r}_i = &-& \sum_{\alpha} {\bf \nabla} V(|{\bf r_i - r_\alpha}|) \nonumber \\
                   &-& \eta \sum_{j < i} {\bf \nabla} V(|{\bf r_j - r_i}|) + {\bf f}_i(T), 
\label{eom}
\end{eqnarray}
where the parameter $\eta$ is used to adjust the relative strength of the adatom interactions with respect to the substrate. 

\begin{figure}[!t]
\includegraphics[width=1.0\columnwidth]{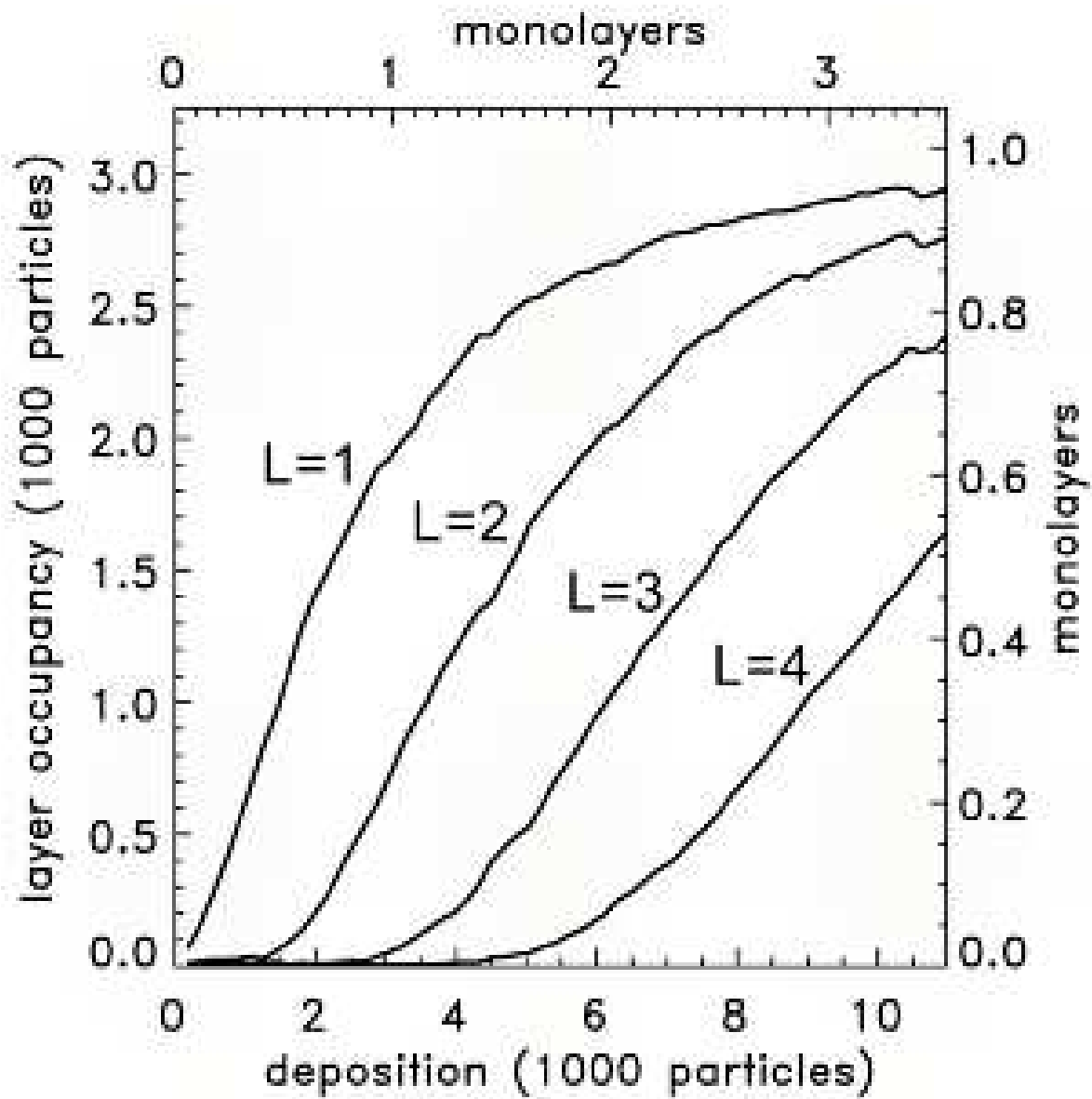} 
\caption{Layer occupancies in layers $L=1-4$ as a function of the total number of adatoms deposited. The top and bottom axes show the total number of atoms deposited in units of the coverage and number of atoms, respectively. The right and left axes are the number of atoms in a given layer, $L=1-4$, in units of the coverage and number of adatoms, respectively. Shown are results  for relative interaction strengths $\eta = 1.2$}
\label{fig:Aldepo}
\end{figure}

In the experiments the quasicrystal sample and its holder serve as a thermal  reservoir that cools down the impinging ``hot" adatoms to maintain the adsorbate and substrate at constant temperature. Since the substrate in our simulation is rigid, it is itself a thermal insulator and we have therefore introduced thermal equilibration artificially by adopting a simple realization of a thermostat with adjustable cooling rate~\cite{Gilmore}. The excess kinetic energy of the adatoms is dissipated by an additional friction force ${\bf f}_i(T)$ that is given by the product of a damping constant $\gamma$ times the excess kinetic energy with respect to the desired average kinetic energy. Further details of the numerical implementation have been described elsewhere \cite{mungan}.

Specifically, we chose a lower damping rate of $\gamma = 0.0625$ (in units, in which the mass of the atoms are set to unity), and a substrate temperature of $kT/\epsilon = 4$, yielding roughly a fourfold increase in the relaxation times compared to those in our earlier work \cite{mungan}. We accordingly increase the waiting times in order to ensure that the substrate and adatoms have relaxed to their equilibrium temperature before a new batch of particles is injected. Note that the damping rate $\gamma$, by establishing the thermal relaxation time for impinging adatoms, also determines the spatial extent to which the adatoms can diffuse across the substrate before losing their excess kinetic energy and getting stuck at an energetically favorable site. The relative strength of the adatom interaction $\eta$, on the other hand, determines how strongly already deposited adatoms can steer this diffusive process. Large values of $\eta$ will cause arriving atoms to be directed towards already deposited adatoms. Our numerical simulations do indeed confirm this picture: For low values of $\gamma$ and $\eta$, {\it e.g.}, such as $\gamma$ = 0.025 and $\eta = 1$ used earlier \cite{mungan}, the injected particles hit and essentially stick to the substrate with relatively little lateral diffusion. 
Under such conditions an initially homgeneous lateral distribution of injected adatoms will give rise to a layer-by-layer type of growth, while keeping $\gamma$ constant and increasing $\eta$ will increase the probability of arriving adatoms to be deposited on already present ones giving rise to a cluster-type growth. 
Both of these growth modes were indeed observed in our simulations \cite{mungan}. In the numerical work presented here, a low value of $\eta = 1.2$ is used along with a much smaller damping rate ($\gamma = 0.0625$). Thus, compared to the simulations mentioned above, impinging adatoms can diffuse longer around the substrate surface. Moreover, with increasing coverage, the accessible favorable binding sites will include both already deposited adatoms as well as substrate sites that have remained exposed.  

We start by investigating the occupancy of the adlayers as a function of the total number of adatoms deposited onto the substrate for $\eta = 1.2$.  Individual layers were determined from the height distributions of the adatoms. In Fig.~\ref{fig:Aldepo}  each continuous curve represents the evolution of a single layer population, as adatoms are injected upon the substrate. Note that even after 3.5\,ML of deposition, the first layer is not yet completed. This implies that the quasicrystalline substrate is only partially concealed at this coverage. 

For coverages up to about 0.5\,ML, the deposition curve of the first layer is rather steep and flattens thereafter.  At these coverages the other layers have not formed yet, and a sticking coefficient of 
about 0.8 can be readily inferred from Fig.~\ref{fig:Aldepo}. Beyond 1.5\,ML all four layers seem to grow simultaneously, implying a cluster type growth. 

\begin{figure}
\includegraphics[width=1.0\columnwidth]{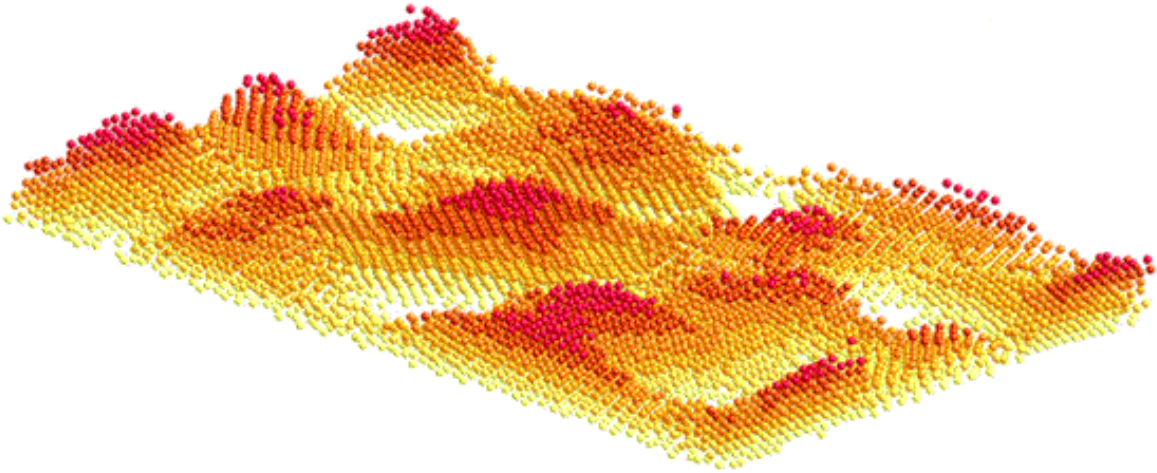}
\caption{(color online) Overall adatom configuration at 3.5\,ML coverage for $\eta = 1.2$. Atoms belonging to a particular layer have the same color. The interface layer is yellow. The adlayer structure, while well-ordered, is highly corrugated and does not fully cover the substrate. Observe the different orientation of atom rows in the domains.}
\label{fig:adconf}
\end{figure}

We now turn to the real-space configuration of the film. Fig.~\ref{fig:adconf} presents the overall atomic configuration of 3.5\,ML film thickness for $\eta = 1.2$. It is  evident that the film is highly corrugated with well-formed hexagonal domain structures. Notice that the adlayer does not fully cover the quasicrystalline substrate. This result is consistent with the deposition curve presented in Fig.~\ref{fig:Aldepo}.  In the experiments the same conclusion was drawn by observing the development of the LEED pattern from the quasicrystalline substrate as a function of Al adsorbate deposition.  

\begin{figure}
\includegraphics[width=0.5\columnwidth]{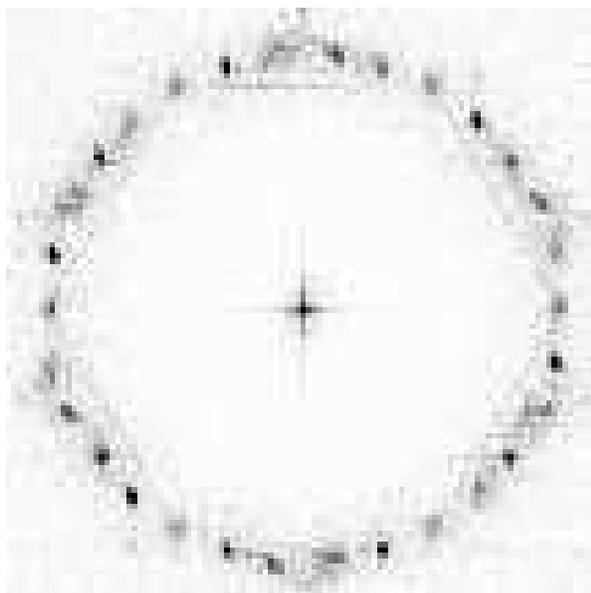}
\caption{Intensity pattern obtained from the Fourier transformation of the adatom structure of the fourth layer. The 30 spots can be grouped into five families of six diffraction spots.}
\label{fig:aldiff}
\end{figure}

In order to further highlight the presence of the multiple domain structure, we present in Fig.~\ref{fig:aldiff} the intensity pattern, as obtained from the Fourier transformation of the fourth adlayer. We clearly distinguish 5 families of six patches, each representing one hexagonal domain orientation of adatoms one of which is highlighted as an aid to the eye. The diffraction pattern consists of rather large patches due to the limited domain size in the fourth layer of the surface constellation, presented in Fig.~\ref{fig:adconf}, having a coverage of about 0.5\,ML ({\it cf}. Fig.~\ref{fig:Aldepo}).

In summary, domain formation can occur at lower relative adatom-adatom interaction strength $\eta$ with increasing thermal relaxation times. A lower damping constant increases the relaxation times for adatoms and permits thereby the exploration of a larger set of energetically favorable local configurations. Our numerical work presented here focused on adsorption at high coverages. However, there are also interesting 
experimental results for adsorption at low coverage, where the diffusion of the adatoms on the 
substrate can lead to the formation of small islands, such as the 5-atom starfish ensembles 
found in icosahedral AlCuFe \cite{Caietal}. The formation of such structures by surface diffusion is rather 
interesting, and has been modelled in terms of a two-dimensional lattice bond gas on a 
disordered bond network that takes into account the basic processes for the occupation and vaccation of 
energetically favorable discrete sites by thermally activated hopping \cite{LatticeBondGas}. It would 
be interesting to compare the key assumption of such models with the kind of three-dimensional simulations of 
adsorption described here. This is left for future work.

\subsection {Oxygen Adsorption}

The reaction of oxygen with crystalline Al is complicated, because each crystallographic surface shows a different characteristic behavior. Upon oxygen exposure, a thin layer of disordered aluminum oxide  is formed on the Al(100) face, while the reaction of oxygen with the Al(111) surface proceeds via a chemisorbed precursor state  \cite{Flod}. LEED studies show that the clean and oxygen-covered surface display the same symmetry. For low defect-density surfaces, the sticking coefficient of oxygen is found to be extremely low and independent of temperature \cite{Zhukov}. It has been argued that at elevated temperatures adsorbed oxygen atoms efficiently migrate on the surface to nucleate in hexagonal islands \cite{Trost}. At high coverages, aluminum oxide grows as a thin amorphous layer that protects the bulk from further oxidation.

The oxidation behavior of the pentagonal surface of the icosahedral Al-Pd-Mn quasicrystal has been reported by several groups. In the earlier studies it was found that  predominantly Al atoms oxidize and form an amorphous alumina layer at the surface \cite {Chang, Popovic}. The chemisorbed phase, which serves as a precursor to oxidation of Al, was found to destroy the quasicrystalline order of the surface \cite {Chang}. As long as the surface is oxidized in ultra-high vacuum conditions, the quasicrystal is found to form a skin of pure aluminum oxide which is about 5 \AA\ thick. Recently, Longchamp and coworkers have shown that high-temperature exposure of the pentagonal surface leads to a well-ordered, atomically thin aluminum oxide layer \cite{Longchamp1}. Similarly, a research team in Nancy, France, studied the kinetics of oxygen chemisorption and oxide growth on the icosahedral Al$_{62}$Cu$_{25.5}$Fe$_{12.5}$ as a function of temperature and oxygen exposure \cite{Nancy}. No account was taken of the atomic structure at the surface.

Here, we observe the adsorption of oxygen onto the tenfold-symmetry surface of a decagonal Al$_{70}$Co$_{15}$Ni$_{15}$ quasicrystal during the initial stage of surface oxidation. At this chemisorbed phase, Al is not oxidized and the film is extremely thin as observed for the icosahedral quasicrystals \cite {Chang, Nancy}. Pure oxygen is introduced into the chamber at a partial pressure of 1$\times$10$^{-8}$\,mbar for typically 1000\,s, while the sample is kept at 870\,K. The oxygen take up and the reaction of oxygen with the surface is confirmed by AES. Similar to the  Al(111) surface, oxygen exposure of the aperiodic surface at 870\,K results in the formation of  a stable chemisorbed layer, which has an ordered structure, consisting of 5 different hexagonal domains \cite{mustafa}. Data on AES show that, upon oxygen adsorption, the alloy constituents remain mostly unaffected, and together with the LEED data, they indicate that the oxygen layer is extremely thin.

\begin{figure}[!t]
\includegraphics[width=0.6\columnwidth]{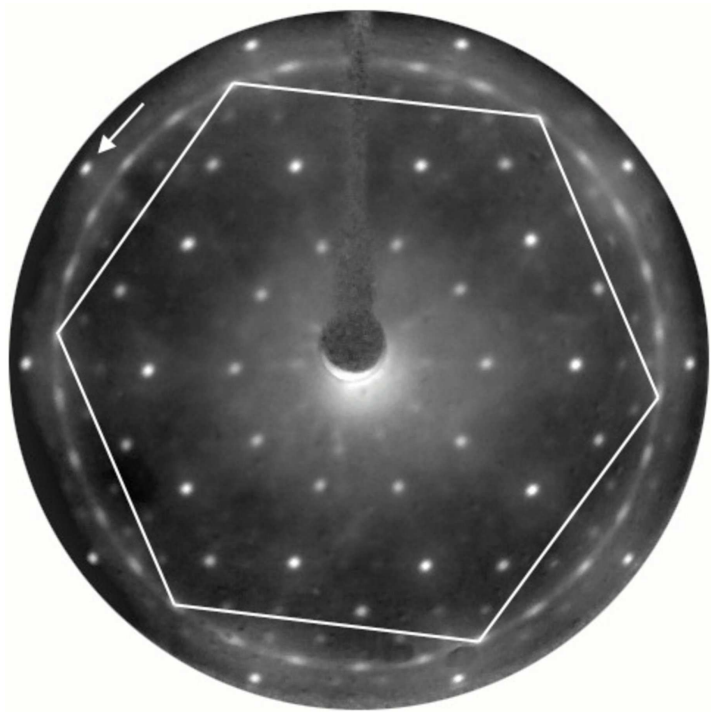} 
\caption{LEED pattern at a primary-electron energy of 50\,eV obtained from the  oxygen-adsorbed tenfold-symmetry surface of decagonal Al-Co-Ni. The pattern is shown after normalization by the response function of the display and recording system in order to eliminate unphysical contributions to the pattern formation. The arrow shows one of the quasicrystal spots. The contribution of oxygen adsorption to the diffraction pattern has a thirtyfold symmetry and, moreover, is in registry with the substrate. The thirtyfold spot pattern can be broken down into five hexagons, one of which is shown.}
\label{fig:Leedox}
\end{figure}

Fig.~\ref{fig:leedonlu}a shows a LEED pattern from the clean tenfold-symmetry surface of Al-Co-Ni at 50\,eV in near-normal electron incidence. The pattern displays tenfold symmetry, while the analysis of the diffraction-spot profiles indicates an average terrace size of at least 100\,\AA, which is in accordance with scanning tunneling microscopy observations \cite{kishida}. The tenfold symmetry in the diffraction pattern has been attributed to a fivefold symmetry in each layer, while two adjacent layers are related by inversion symmetry.

\begin{figure}[!t]
\includegraphics[width=1.0\columnwidth]{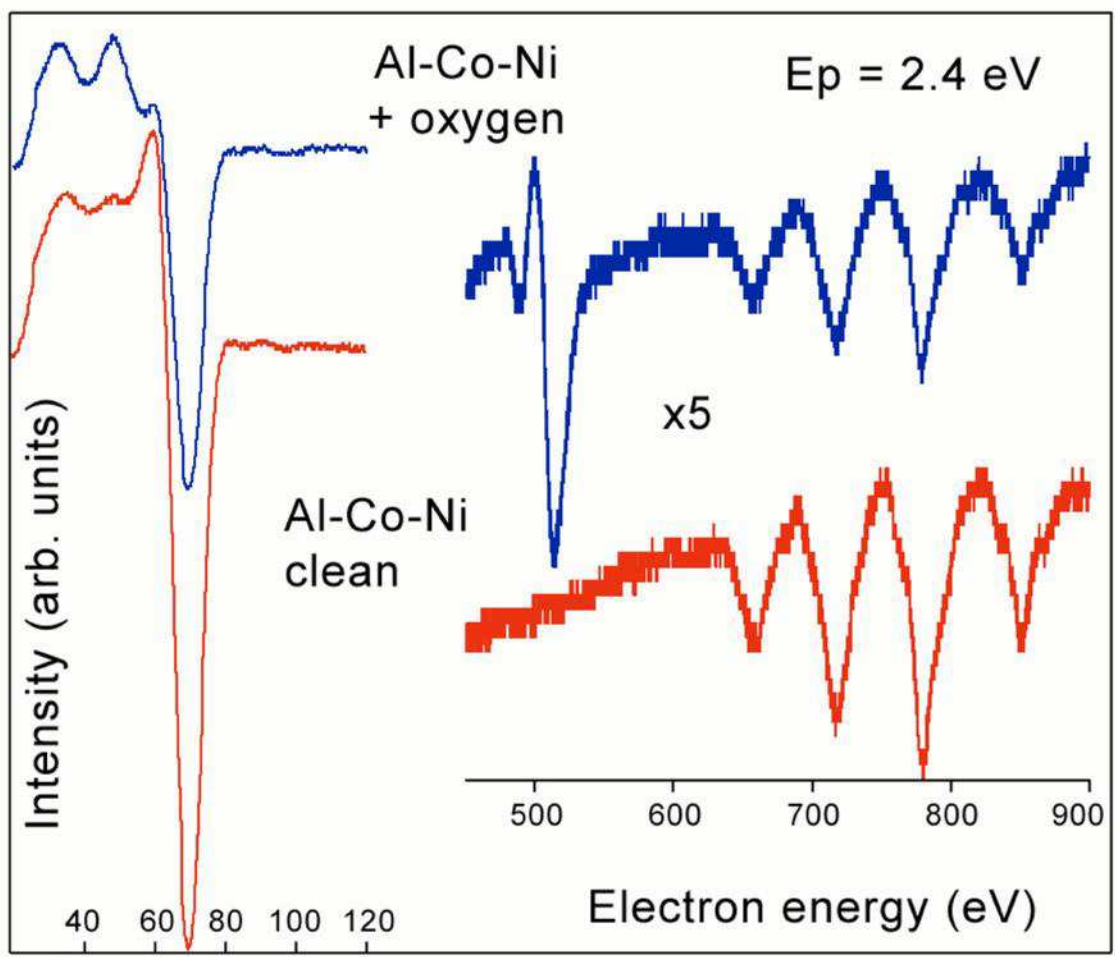} 
\caption{Results of AES applied to the clean (bottom) and oxygen-adsorbed (top) tenfold-symmetry surface of decagonal Al-Co-Ni. Spectra in the region of O (around  500\,eV), Co, and Ni transitions are multiplied by 5.}
\label{fig:aes}
\end{figure}

The LEED pattern depicted in Fig.~\ref{fig:Leedox} is obtained from the same surface after exposure to an oxygen partial pressure of 1\hspace{1pt}$\times$\hspace{1pt}10$^{-8}$\,mbar at 870\,K for 1000\,s. The diffraction spots characteristic of the clean quasicrystalline surface are still clearly visible indicating that the surface film is rather thin and similar to oxide films grown on Al-Pd-Mn at elevated temperatures \cite{Longchamp1,Longchamp2}. The observation of these spots further indicates that the formation of the surface layer preserves the quasicrystalline order at the interface. Moreover, thirty new diffraction features placed evenly on a polar circle of 40.8\,$\pm 1^{\circ}$ emerge. These spots represent the fivefold repetition of a sixfold-symmetric pattern in azimuthal increments of $72^{\circ}$, in accordance with the local symmetry of the quasicrystal surface. One such hexagon is superimposed on the pattern for easy identification. The spots are very narrow in the radial direction, but display appreciable azimuthal smearing. There is no energy-dependent intensity modulation of the spots, and we conclude that they represent a signal from a hexagonal distribution of atoms with an interatomic distance of 3.06\,$\pm 0.05$\,\AA. This value is about 10\,\% larger than the O-O distance in $\alpha$- and $\gamma$-Al$_2$O$_3$. A lattice expansion of a similar amount has also been observed in the oxide films grown on the (110) surface of the cubic binary alloy AlNi \cite{Jaeger,Lykhach}. The limited radial spread of the diffraction spots indicates that the domains are almost as large as the terraces of the quasicrystalline substrate. The azimuthal spread of each of the spots is about 3$^{\circ}$ with nonvanishing intensity inbetween. Similar spreads have been observed for almost all cases of adsorbed atoms and molecules on quasicrystal surfaces, indicating that it is a result of the misfit produced by the aperiodic surface structure at the interface with a periodic structure. However, it is remarkable that the oxygen layer grows on the quasicrystalline substrate in such a large domain size. The azimuthal degree of freedom and the minute thickness of the film may well be the reason that prevents the build up of lattice strain in the oxygen layer.

Fig.~\ref{fig:Leedox} further shows that the oxygen adlayer is in registry with the quasicrystalline substrate in such a way that the quasicrystal spots near the rim of the collector fall inbetween two oxygen spots. It is interesting to note that while a monolayer of Xe or other noble gases form a pseudomorphic layer  \cite{setya}, oxygen adsorption at these coverages already shows a pattern compatible with sixfold symmetry. Note that there is a $6^{\circ}$ azimuthal difference between the locking of the aluminum and oxygen hexagonal structures to the quasicrystal. 

The dissociative adsorption of oxygen has already been observed on Al(111) \cite{Flod}, where the formation of hexagonal oxygen structure is compatible with the substrate symmetry. In the present case, however, the substrate, while containing Al, has a quasicrystalline structure on which the sixfold symmetric oxygen adlayer assembles. While the Al-O interface for Al(111) is stable (nonreactive) up to temperatures of 470\,K, Al in the quasicrystalline Al-Co-Ni matrix remains nonreactive at temperatures close to 900\,K.

Fig.~\ref{fig:aes} presents scans of AES near the Al, O, Ni, and Co transitions for the clean (bottom) and oxygen exposed (top) surfaces. The spectrum of the surface after oxygen adsorption shows no energy shift for Al and transition metal signals, but some reduction of the intensity. This observation indicates that no appreciable electron transfer from a metal site to oxygen has taken place and oxygen is in a chemisorbed state.  In the oxide phase, however, there would be a transfer of $3s^23p$ electrons from Al to O, and there would remain no valence electrons at the Al site to participate in the Auger transition. The $2p$ core-hole state would then relax via an interatomic process emitting an Auger electron with a kinetic energy about 10\,eV lower than the Auger transition energy in Al metal. In the present case, we cannot completely exclude the presence of a minute amount of aluminum oxide, and the  small structure at the low-energy side of the Al transition at the top spectrum may well be a signal for the early stage of oxidation. Yet, we may conclude that the majority of oxygen is in the chemisorbed state and, taking into account the features of the LEED pattern, that the adsorbed oxygen may be just one single monolayer thick. We note that this stage persists at oxygen pressures of 1\hspace{1pt}$\times$\hspace{1pt}10$^{-8}$\,mbar and does not change upon further annealing. The adsorbed oxygen layer acts as an efficient capping on the quasicrystalline surface. The formation of ordered Al oxide layers grown at higher temperatures and oxygen partial pressures is the subject of further work \cite {mustafa}.

\section{Discussion and Conclusions}

We have compared SEI results from the tenfold- and twofold-symmetry surfaces of Al-Co-Ni with SSC calculations based on the structural model presented here. Within the information depth of $20-30$\,\AA\, the details of the experimental features were well reproduced. Electron diffraction from the quasicrystalline surface is also well accounted for by the Fourier transform of the bulk-terminated model surface. The close correspondence between these results provides strong support for the validity of the model used in our calculations.

We have used a quasicrystalline bi-layer perpendicular to the periodic direction of our structural model to investigate the adsorption dynamics of atoms. We were able to reproduce the formation and the size selection of domains as well as their distinct orientations. All these experimental and computational studies, which are extremely sensitive to structural details, furnish a critical test for the quality of the structural model.

Our experimental and computational efforts in determining the decagonal quasicrystal structure as well as the structure of growing adlayers have shown that there is a structural registry between the crystal and the quasicrystal on a local scale.  Experimentally observed diffraction patterns contain an orientational smearing of the structure with a well-defined length scale associated with adatom spacing and symmetry. 

These results indicate that epitaxial aluminum atoms adsorbed on the quasicrystalline surface experience competing interactions. While aluminum atoms favor ordering in their stable fcc phase, the quasicrystalline surface template they condense on, force them into an aperiodic order. The system finds the best compromise by partially satisfying both conditions and breaks up into small aluminum domains ordered in their bulk fcc structure, where each domain remains locally commensurate and in registry with the substrate. This registry results in five distinct orientations for the aluminum fcc nanocrystals. The size of these domains is determined by the interfacial strain energy, as this size increases the island edges get rapidly out of registry with the substrate and it becomes energetically more favorable to break-up into locally commensurate domains.

In heteroepitaxial systems the interfacial strain energy increases with the film thickness, and beyond a critical thickness strain built in the film causes the film to relax to its stable bulk phase by creating misfit dislocations at the interface. In monoatomic systems on decagonal quasicystalline surfaces we studied, this critical thickness does not seem to extend beyond a monolayer thickness, and the system relaxes to small domains of the bulk phase. It is likely, however that, in some carefully chosen binary and ternary alloy systems the critical thickness may increase significantly, and one may be able to grow epitaxy stabilized quasicrystalline phases, that do not exist in nature.

\ack The authors gratefully acknowledge M. G\"undo\u gan and J.-N. Longchamp, who were involved at the initial stages of oxygen-adsorption project. The computations were mainly done using the  Gilgamesh and Kassandra computer clusters at the Feza G\"ursey Institute and Bo\u gazi\c ci University. This work has been funded in part by grant 08B302 of Bo\u gazi\c ci University. Financial support by Schweizerischer Nationalfonds is greatfully appreciated.

\end{document}